\documentclass[%
 twocolumn,
 superscriptaddress,
 amsmath,amssymb,
 aps,
 prl,
]{revtex4-1}

\usepackage{hyperref}

\usepackage[dvips]{graphicx} 
\usepackage{bm}
\usepackage{float}
\usepackage{todonotes}

\graphicspath{{./}{../Figures//}{../Figures/}}
\DeclareGraphicsExtensions{.eps}

\newcommand {\be}{\begin{eqnarray}}
\newcommand {\ee}{\end{eqnarray}}

\begin{document}


\title{Rotated stripe order and its competition with superconductivity in La$_{1.88}$Sr$_{0.12}$CuO$_4$}

\author{V. Thampy}
\author{M. P. M. Dean}
\affiliation{Condensed Matter Physics and Materials Science Department, Brookhaven National Laboratory, Upton, New York 11973, USA}

\author{N. B. Christensen}
\affiliation{Department of Physics, Technical University of Denmark (DTU), DK-2800 Kongens Lyngby, Denmark}

\author{Z. Islam}
\affiliation{The Advanced Photon Source, Argonne National Laboratory, Argonne, Illinois 60439, USA}

\author{M. Oda}
\affiliation{Department of Physics, Hokkaido University, Sapporo 060-0810, Japan}
\author{M. Ido}
\affiliation{Department of Physics, Hokkaido University, Sapporo 060-0810, Japan}
\author{N. Momono}
\affiliation{Department of Materials Science and Engineering, Muroran Institute of Technology, Muroran, Japan}

\author{S. B. Wilkins}
\author{J. P. Hill}
\affiliation{Condensed Matter Physics and Materials Science Department, Brookhaven National Laboratory, Upton, New York 11973, USA}

\date{\today}

\begin{abstract}
We report the observation of a bulk charge modulation in La$_{1.88}$Sr$_{0.12}$CuO$_4$ (LSCO)  with a characteristic in-plane wave-vector of (0.236, $\pm \delta$), with $\delta$=0.011 r.l.u. The transverse shift of the ordering wave-vector indicates the presence of rotated charge-stripe ordering, demonstrating that the charge ordering is not pinned to the Cu-O bond direction. On cooling through the superconducting transition, we find an abrupt change in the growth of the charge correlations and a suppression of the charge order parameter indicating competition between the two orderings. Orthorhombic LSCO thus helps bridge the apparent disparities between the behavior previously observed in the tetragonal ``214'' cuprates and the orthorhombic yttrium and bismuth-based cuprates and thus lends strong support to the idea that there is a common motif to charge order in all cuprate families.
\end{abstract}

\maketitle

Charges doped into the copper oxide planes of the insulating parent compounds frequently organize into rich electronic textures. The most well-known example is the charge-spin stripe state, first seen in La$_{1.48}$Nd$_{0.4}$Sr$_{0.12}$CuO$_4$ ~\cite{Tranquada1995}, and subsequently in a number of other ``214'' cuprates, most notably La$_{2-x}$Ba$_x$CuO$_4$ (LBCO) ~\cite{PhysRevB.70.104517, PhysRevB.78.174529, Abbamonte2005, Wilkins2011, Hucker2012, DeanLBCO2013, PhysRevB.83.104506}. More recently, charge modulations have also been observed in other cuprate families including YBa$_2$Cu$_3$O$_{6+y}$ (``123'') ~\cite{Ghiringhelli17082012, Chang2012a, Achkar2012, PhysRevLett.110.187001, PhysRevLett.110.137004, LeTacon:2013es, Marc.Nat.477.2011, BlackburnYBCOIXS2013}, and (Bi$_{2-x}$Pb$_x$)(Sr$_{2-y}$La$_y$)Ca$_{n-1}$Cu$_{n}$O$_{2n+4+\delta}$ ~\cite{Comin:2014ck, Comin:2014vq, Rosen:2013ef, Neto2014, Hashimoto2014}. This suggests that the charge ordering (CO) in the copper oxide planes stems from a fundamental underlying instability common across the different cuprate families.

However, this apparent ubiquity belies significant differences in the CO characteristics between the different families. In the 214 compounds, the charge order occurs at a wavevector $q_{CO}\sim$~$0.24$ r.l.u. and is accompanied by static spin order of twice the period, leading to a picture of charge stripes which form antiphase domain walls separating regions of antiferromagnetic order. These are stabilized by the low temperature tetragonal (LTT) phase present in these materials ~\cite{Fujita:2002ia, Fujita:2002by, Kampf:2001ib}, leading to a stripe phase that is locally commensurate ~\cite{Robertson:2006bc}. At x=1/8, these LTT stabilized charge spin stripes strongly suppress bulk superconductivity, e.g. in LBCO, T$_C$ $< 2$~K~ \cite{Moodenbaugh:1988bm, PhysRevLett.99.067001}.

In contrast, in the orthorhombic yttrium- and bismuth- based compounds, which lack the LTT phase, the charge order occurs at $q_{CO}\sim$~$0.3$ r.l.u.\ and there is no static spin order down to at least 5 K ~\cite{Haug:2009io, Stock:2002hn, Dai:2001ej, Enoki:2013fd}. Instead, the magnetism is dynamic and though it is incommensurate, the spin fluctuations do not occur at twice the period of the charge order \cite{PhysRevLett.110.137004}. Indeed, the spin wave vector has the opposite doping dependence to that of the charge order wave vector ~\cite{PhysRevLett.110.137004}. In these systems, a Fermi-surface nesting picture has been invoked to describe the origin of the charge order ~\cite{Ghiringhelli17082012, Comin:2014ck, Comin:2014vq, Rosen:2013ef}. While the charge modulations in the 123 family do not suppress superconductivity  as strongly as in LBCO, there is clear evidence that they are coupled to superconductivity: specifically, the intensity and correlation length of the charge modulation both peak at $T_c$ ~\cite{Ghiringhelli17082012, Chang2012a, PhysRevLett.110.187001, PhysRevLett.110.137004}.

Reconciling the differences between the two classes is essential to develop, or rule out, a unified picture of electronic ordering in the cuprates, and to understand the relationship between the electronic order, superconductivity and the pseudogap phase. To address this, what is required is something of a ``missing link'' compound between the two families. That is an orthorhombic compound without an LTT crystal structure that has spin and charge stripe order.

Here we report that La$_{1.88}$Sr$_{0.12}$CuO$_4$ (LSCO) is just such a compound. It exhibits bulk charge order at (0.236(4), $\pm \delta$), with $\delta$=0.011(1) r.l.u.\ The ordering wave-vector is close to that observed in the other 214 compounds, but is shifted in the transverse direction, demonstrating the charge stripes, as well as the spin stripes ~\cite{Kimura:2000ux}, are rotated by $\sim$~3$^{\circ}$ away from the Cu-O bond direction, and consequently not locked to the high symmetry directions in the lattice. On entering the superconducting phase, the growth of the charge correlations is interrupted, indicating that both phenomena compete for the same electrons. Taken together with existing results in 123 and 2212 cuprates, these new results suggest that charge ordering in the different cuprates has a common phenomenology, and interacts with bulk superconductivity in similar ways despite differences in for example, wave-vector and spin order. In particular, it suggests that charge order and superconductivity are delicately balanced in these materials. If the charge order is sufficiently well correlated and/or pinned to the lattice, as in LBCO, it can prevent the formation of a coherent superconducting state \cite{PhysRevLett.99.067001}. A weakly correlated charge modulation ($\xi_{\mathrm{CO}} \lesssim $ 100 \AA) not pinned to the lattice, as is the case in LSCO examined here, allows superconducting phase coherence and consequently a bulk superconducting state to develop at the cost of the charge order parameter.

The LSCO sample used for this experiment was a single crystal grown using the floating zone method ~\cite{Nakano:1998vz} and cleaved {\em ex situ}, to reveal a [001] surface normal. At room temperature, its crystal structure is tetragonal with space group ($I4/mmm$) and lattice parameters $a = b = 3.78$~\AA, $c = 13.23$~\AA~. Below about 255 K, a structural transition to the  $Bmab$ space group ~\cite{RADAELLI:1994ua} occurs forming twinned orthorhombic domains. Despite this, throughout this paper  we will index reciprocal space using the high temperature tetragonal (HTT) unit cell, for ease of comparison with other studies. The sample has a hole concentration ($x \approx 0.12$) ~\cite{Nakano:1998vz}, i.e., close to the doping for which there is a plateau in the $x$-$T_c$ phase diagram \cite{PhysRevB.73.180505}.

The soft x-ray diffraction experiments were carried out on the X1A2 beamline at the National Synchrotron Light Source (NSLS), Brookhaven National Laboratory, using photons with energies at the peak of the the Cu $L_3$ x-ray absorption spectrum ($2p_{3/2} \rightarrow 3d$), which enhances the sensitivity to lattice distortion caused by charge ordering \cite{PhysRevB.74.195113, FinkRepProgPhys2011}.  The sample orientation (UB matrix) was determined using the (002) and (101) Bragg reflections, and a CCD detector was used to collect the scattered intensity, which was then rebinned to obtain two-dimensional (2D) slices through reciprocal space. We note that the scattered intensities are not energy resolved, and have a substantial contribution from inelastic scattering. This contribution is, however, only weakly dependent on $q$ and was subtracted as a flat background \cite{NPh000294485400021, Dean:2013cy, PhysRevB.88.020403, PhysRevLett.92.117406, Ghiringhelli17082012}.

Hard x-ray diffraction experiments on the same sample were conducted on beamlines X22C at the NSLS and 6ID-B at the Advanced Photon Source (APS), Argonne National Laboratory. In each case, an incident photon energy of 8.9 keV was chosen to avoid the fluorescence background from copper emission. The scattered x-rays were detected using a point detector. Both the hard and soft x-ray measurements were conducted in a vertical scattering geometry, with the [001] and [100] directions lying in the scattering plane, and $\sigma$ polarized x-rays.

\begin{figure}[t]
    \includegraphics*[trim = 0mm 0mm 0mm 0mm, clip,width=.99\columnwidth]{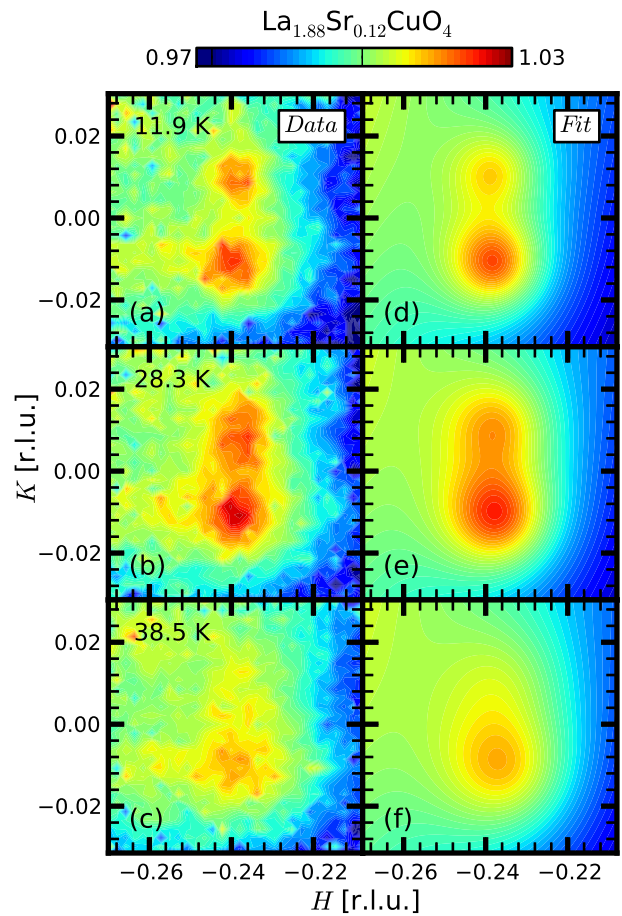}
	\caption{(Color online) CO scattering intensity in the $(H, K, 1.5)_{HTT}$ plane collected at photon energies set to the peak in the Cu $L_3$-edge absorption, and integrated over $1.48 \le L \le 1.52$ for (a) 12 K, (b) 28.3 K and (c) 38.5 K. The data are normalized to the background intensity for each temperature. (d-f) Results of corresponding 2D fits to two Lorentzian-squared functions with a planar background.}
\label{fig:Slices}
\end{figure}

\begin{figure}[t]
    \includegraphics*[trim = 0mm 0mm 0mm 0mm, clip,width=0.99\columnwidth]{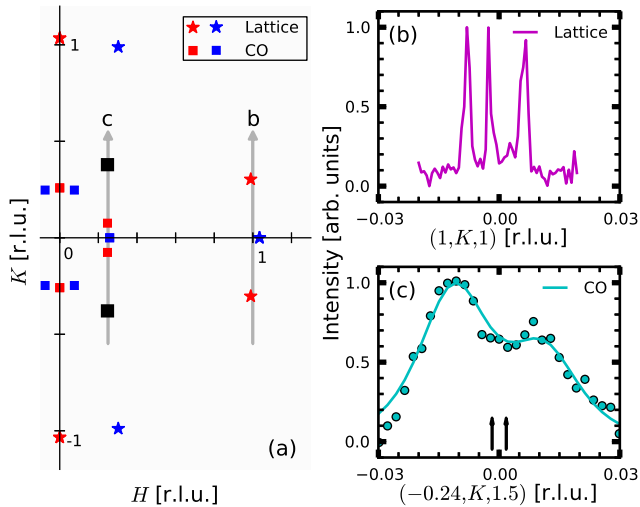}
	\caption{(Color online) (a) Schematic drawing showing the peak positions of the structural lattice and CO Bragg peaks in reciprocal space arising from the four possible domains in twinned LSCO (see text). The magnitude of the orthorhombic distortion and CO incommensurability are exaggerated by a factor of 40 for clarity. The red and blue squares show the calculated positions of the CO peaks corresponding to the individual domains if there were no incommensurability in the transverse direction. The filled black squares show the positions of the observed peaks. The arrows show the direction of $K$ scans through the $(1,0,1)$ fundamental Bragg peak (b) and $(-0.236,0,1.5)$ CO peak (c). The separation between the CO peaks is significantly larger than that between the structural peaks even though the magnitude of the in-plane wave-vector is four times smaller. The arrows in (c) show the expected positions of the CO peaks based on orthorhombic splitting.}
    \label{fig:Ortho}
\end{figure}

\begin{figure}[t]
    \includegraphics*[trim = 0mm 0mm 0mm 0mm, clip,width=0.99\columnwidth]{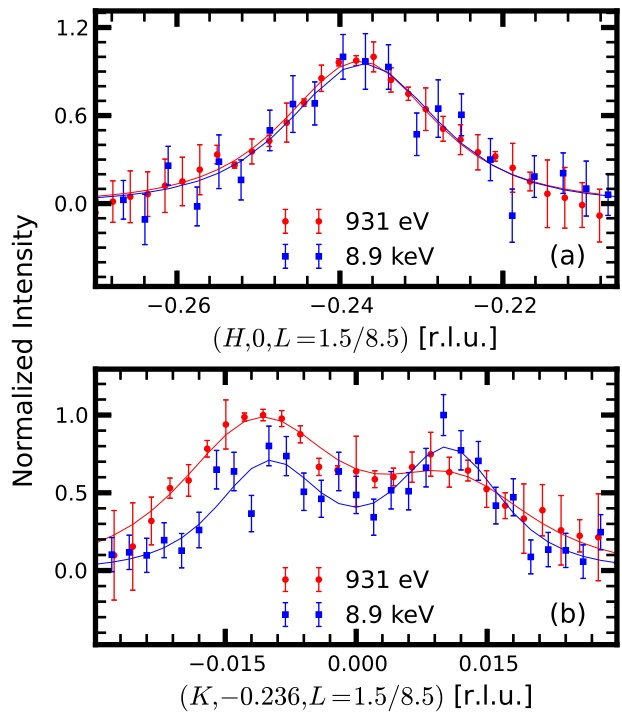}
	\caption{(Color online) Background subtracted hard x-ray data plotted along with soft x-ray data collected at the Cu $L_3$-edge. (a) Scans along $H$ through the CO peak at $L=1.5$ at 931 eV (shown in red), and at $L=8.5$ at 8.9 keV (shown in blue). (b) $K$ scans through the same peaks showing the splitting along the transverse direction.}
    \label{fig:6ID}
\end{figure}

\begin{figure}[t]
    \includegraphics*[trim = 0mm 0mm 0mm 0mm, clip,width=.99\columnwidth]{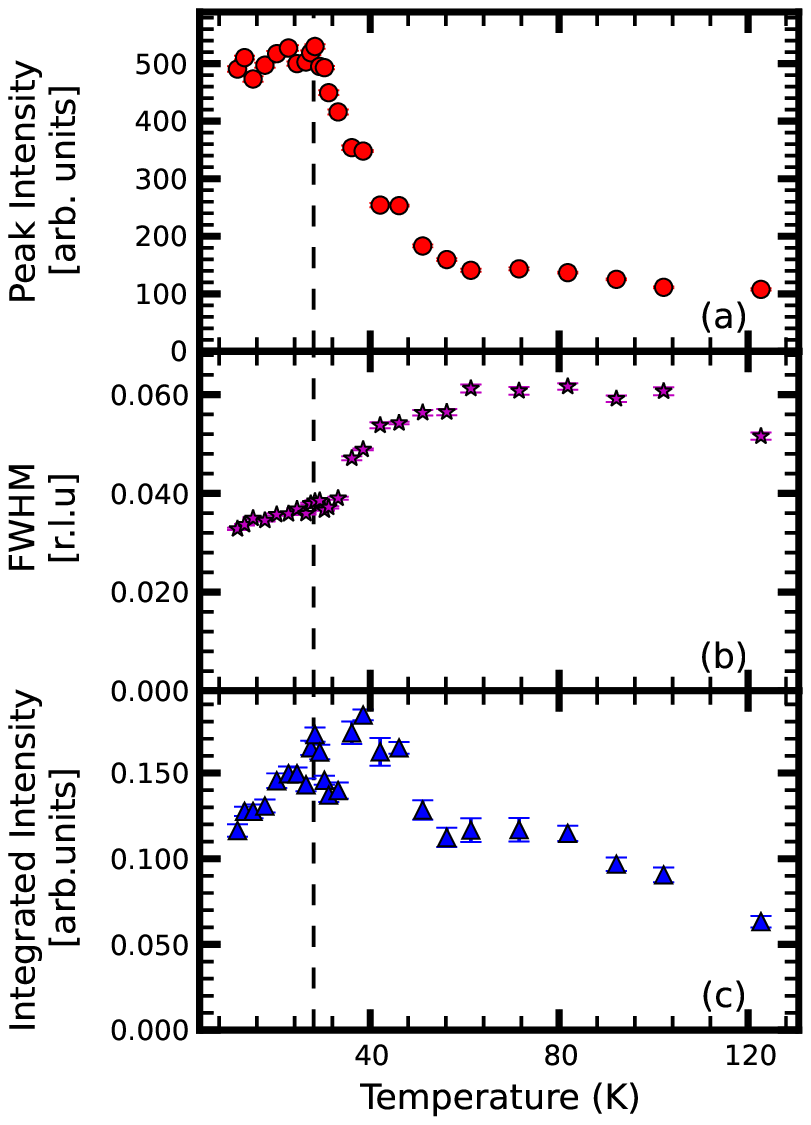}
	\caption{(Color online) Temperature dependece of the (a) Peak intensity, (b) FWHM, and (c) Integrated intensity, of the scattering from the CO at the Cu $L_3$-edge. The dotted line shows the superconducting transition, $T_{\mathrm{C}}$. The peak intensity and the integrated intensity show a maximum at $T_{\mathrm{C}}$. At the same time, the correlations, which are inversely proportional to the FWHM, develop and grow stronger below $\sim$~60 K, but are suppressed on entering the superconducting state.}
\label{fig:TDep}
\end{figure}

Figures~\ref{fig:Slices}(a-c) show the momentum dependence of the scattering in the $HK$ plane, integrated over $1.48 \le L \le 1.52$, at $T$ = 11.9 K, 28.3 K, and 38.5 K, as measured at the Cu L$_3$ edge..  Two peaks are observed at H=0.236(4) r.l.u., split in the transverse, $K$ direction. Given the orthorhombic crystal structure, it is not surprising to see multiple peaks arising from twin domains rotated with respect to each other~\cite{Wakimoto:2006fq}. However, the angle of rotation between the twins ($\sim$~0.3$^{\circ}$~\cite{Wakimoto:2006fq}) cannot account for the observed positions of the peaks, as illustrated in Figure~\ref{fig:Ortho}. In the LTO phase, the LSCO system generally includes four possible twins ~\cite{Wakimoto:2006fq} comprising two mirror pairs, rotated by 90 degrees with respect to each other. The Bragg peaks arising from the splitting of the $(1,0,1)_{HTT}$ reflections for the two pairs of domains are shown in red and blue stars respectively in Figure~\ref{fig:Ortho}(a). The red and blue squares show the corresponding calculated locations for CO peaks, assuming the CO superlattice has the same symmetry as the underlying structural lattice. The expected orthorhombic splitting for the CO is $\delta=\pm 0.0017$ r.l.u.. In fact, as shown in Figure~\ref{fig:Ortho}(b), the CO peaks have $\delta=\pm 0.011$ r.l.u. -- as depicted by the filled black squares in Figure~\ref{fig:Ortho}(a). In Figure~\ref{fig:Ortho}(b) transverse ($K$) scans through the (1,0,1) Bragg peaks are shown, here the splitting is precisely the expected value of 0.007 r.l.u. at this Q. Thus we conclude that the two peaks around (0.24,0) do not arise from orthorhombic splitting, but rather are due to an intrinsic transverse incommensurability of the charge order itself.

Previous works have reported a transverse incommensurability for elastic {\em magnetic} peaks in La$_{1.88}$Sr$_{0.12}$CuO$_4$ ~\cite{Kimura:2000ux, Romer:2013es} and in La$_2$CuO$_{4+\delta}$ ~\cite{Lee:1999vk}. This shift was described by the angle of rotation ($\theta_Y$) of the spin density modulation direction away from the tetragonal axes. The angle was $\sim$~$3^{\circ}$ in both cases. Here we report that the CO wave-vector is rotated by $\sim$~$2.7^{\circ}$, which is comparable to that of the magnetic peaks. It seems natural then to conclude that the magnetic and charge peaks arise from a single coherent charge and spin density wave structure  (stripes) that is rotated by $\sim$~$3^{\circ}$ from the Cu-O bond direction. This is the first such observation of rotated charge stripes. Rotated stripes are consistent with the predictions from a Ginzburg-Landau analysis of the stripe order parameter ~\cite{Robertson:2006bc}.

Since the tetragonal symmetry of the crystal structure is already lost when it goes through the LTO transition at $\sim$~$255$~K, there is no \emph{a priori} reason to expect the CO to exhibit tetragonal symmetry. However, orthorhombicity is not a sufficient condition to explain the $Y$ shift since no such shift is seen in YBCO, which is also orthorhombic. According to the analysis in  ~\cite{Robertson:2006bc}, the key to the rotated stripe order is the {\em rhombohedral} distortion of the Cu-O plaquette in LSCO. One way rotated stripes can be accommodated is through kinks along the charge walls, where the average separation between the walls changes from three spins to four spins, to account for the deviation of $q$ from 0.25. A tilt angle of $\sim$~$2.7^{\circ}$ would correspond to a kink roughly every 21 Cu sites along a stripe. Measuring the doping dependence of CO could shed light on the appropriateness of this picture ~\cite{Bosch:2001dn}. Finally, we note that the CO sets in at a higher temperature than the spin order ($T_{SO}$=30 K ~\cite{Kimura:2000ux}), suggesting that the CO energetics sets the stage for the rotated geometry of the spin charge stripes.

Given that previous work suggested that charge ordering in LSCO is a surface phenomenon ~\cite{Wu:2012gh}, we utilized hard x-ray scattering to probe deep (several microns) into the sample. Figure~\ref{fig:6ID} compares $H$ and $K$ scans through the charge order taken at 8.9 keV (blue squares) and 931 eV (red circles). The data lie on top of each other, and yield the same correlations lengths. Furthermore, we observe a transverse split of the same magnitude with hard x-rays as was seen with soft x-rays. As demonstrated recently in Ref.~~\cite{Christensen}, we conclude that the charge order in LSCO is a bulk phenomenon. We also measured the integrated intensity of the charge order in La$_{1.875}$Ba$_{0.125}$CuO$_4$ using hard and soft x-rays under the same experimental conditions as for the present data. At both energies, the LSCO integrated intensity is found to be $\sim$~4 times weaker than La$_{1.875}$Ba$_{0.125}$CuO$_4$ \cite{Thampy:2013cl}.

To see the effect of superconductivity on the charge order, we next look in detail at the temperature dependence of the scattering. To do so, we fit a two dimensional function comprising two isotropic Lorentzian-squared functions of equal width and centered at ($H_o, \pm K_o$) on a plane background to the data such as shown in  Figures~\ref{fig:Slices}(a-c). The fits yield peak positions of $(-0.236(4), \pm 0.011(1))$. This value is found to be independent of temperature, and was therefore held fixed for all fits. Example resulting fits are shown in Figures~\ref{fig:Slices}(d-f) along side the respective experimental data. Figure~\ref{fig:TDep} shows the evolution of the peak intensity, the full-width-at-half-maximum (FWHM), and integrated intensity of the scattering as a function of temperature. Significant charge order sets in at $T_{CO} \sim$~$60$~K. Above this temperature, it is hard to distinguish a clear peak, although there are indications of remnant intensity, which suggests that correlations might persist at higher temperatures. Below 60 K, the peak intensity rises sharply, accompanied by a corresponding decrease in the FWHM. The integrated intensity also rises below $T_{\text{CO}}$. An abrupt change is seen in all these parameters as the superconducting state is entered. This is most obviously seen for the peak intensity, which shows a clear suppression below $T_c$. The FWHM continues to decrease, though the rate of decrease is much slower than above $T_c$. The correlation length, calculated as ($\frac{1}{\mathrm{HWHM}}$), increases slightly from $\sim$~$55$~\AA~ at $T_c$ to $\sim$~$60$~\AA~ at $T=11.3$~K. The order parameter, as measured by the integrated intensity, decreases below $T_c$.

The increasing correlation length below T$_C$ is suggestive of microscopic co-existence where CO exists throughout the bulk  and not just in phase segregated regions. Though our data does not rule out the possibility that there could be regions of LTT phase ~\cite{Bozin:1999wi, Christensen} where the CO correlations continue to grow in the SC state, even as they are suppressed in the rest of the volume.

The thermal evolution of the parameters characterizing the charge order is reminiscent of that seen in YBCO ~\cite{Chang2012a, Ghiringhelli17082012}, and indicates competition between the superconducting and CO order parameters. One commonality is the short range of the correlations in both the materials, $\sim$~$55$~\AA~ in YBCO, and $\sim$~$65$~\AA~ in LSCO. Evidently, the shorter correlations, coupled with the relatively higher superconducting transitions temperature as compared to LBCO, do not allow the charge order to develop fully. There is also  the possibility that when charge stripe order is pinned more strongly to the lattice in the LTT phase, it is more disruptive to inter-planar SC coherence, and consequently suppresses SC more strongly. 

To conclude, we have observed charge ordering in LSCO with the characteristic in-plane wave-vector rotated away from the crystal axes direction by $\sim$~2.7$^{\circ}$. The concomitant rotation of the elastic magnetic peaks ~\cite{Kimura:2000ux} evinces a unique rotated charge spin stripe order hitherto unseen in other cuprates. Whereas the off-axes wave-vector sets LSCO apart, the thermal evolution of the parameters characterizing the charge order and its antagonistic coupling to superconductivity puts it firmly on the same footing as the other cuprates. This competition between the charge order and superconductivity has been seen most clearly in the yttrium and bismuth-based cuprates which do not show any static magnetic order. Our results clearly demonstrate that charge and spin stripe order, which so far have only been observed in the 214 family, vies with superconductivity in much the same way, suggesting a common motif of intertwined electronic degrees of freedom possibly arising from the same multi-component order parameter ~\cite{Tsvelik:2014wx}.

We would like to thank John Tranquada, Akash Maharaj, Wei Ku and Weiguo Yin for helpful discussions. Work performed at Brookhaven National Laboratory was supported by the U.S. Department of Energy, Division of Materials Science, under Contract No. DE- AC02-98CH10886. Use of the Advanced Photon Source, an Office of Science User Facility operated for the U.S. Department of Energy (DOE) Office of Science by Argonne National Laboratory, was supported by the U.S. DOE under Contract No. DE-AC02-06CH11357. Use of the National Synchrotron Light Source, Brookhaven National Laboratory, was supported by the U.S. Department of Energy, Office of Science, Office of Basic Energy Sciences, under Contract No. DE-AC02-98CH10886. This work was supported the Danish Agency for Science, Technology, and Innovation under DANSCATT.

\bibliography{LSCO}

\end{document}